\documentstyle[a4,11pt]{article}
\newcommand{\beq}{\begin{equation}}
\newcommand{\eeq}{\end{equation}}
\newcommand{\bea}{\begin{eqnarray}}
\newcommand{\eea}{\end{eqnarray}}
\newcommand{\rf}[1]{(\ref{#1})}

\newcommand{\pa}{\partial}

\renewcommand{\L}{\Lambda} 
\renewcommand{\b}{\beta}

\newcommand{\k}{\kappa}

\newcommand{\oh}{\frac{1}{2}}

\newcommand{\equ}{\!=\!}

\newcommand{\tL}{{\tilde{\L}}}

\newcommand{\tY}{{\tilde{Y}}}
\newcommand{\tZ}{{\tilde{Z}}}
\newcommand{\ty}{{\tilde{y}}}
\newcommand{\tz}{{\tilde{z}}}
\newcommand{\tg}{{\tilde{g}}}
\newcommand{\tG}{{\tilde{G}}}

\newcommand{\tT}{{\tilde{T}}}

\newcommand{\SL}{{\sqrt{\L}}}
\newcommand{\tSL}{\sqrt{\tL}}
\newcommand{\FL}{\L^{1/4}}
\newcommand{\bZ}{{\bar{Z}}}

\begin{document}
\input{epsf}
\topmargin 0pt
\oddsidemargin 5mm
\headheight 0pt
\headsep 0pt
\topskip 9mm
\pagestyle{empty}

\hfill NBI-HE-99-50

\hfill AEI-1999-51

\addtolength{\baselineskip}{0.20\baselineskip}

\begin{center}
\vspace{26pt}
{\large \bf
On the relation between Euclidean and Lorentzian 2D quantum gravity  }

\vspace{26pt}

\vspace{18pt}
{\sl J.\ Ambj\o rn, J.\ Correia and C.\ Kristjansen}\hspace{0.025cm}\footnote
{E-mail: ambjorn, correia, kristjan@nbi.dk}
\\
\vspace{6pt}
The Niels Bohr Institute \\
 Blegdamsvej 17,
DK-2100 Copenhagen \O, Denmark \\

\vspace{18pt}
{\sl R.\ Loll}\hspace{0.025cm}\footnote{E-mail:
loll@aei-potsdam.mpg.de} \\
\vspace{6pt}
Max-Planck-Institut f\"ur Gravitationsphysik
(Albert-Einstein-Institut), \\
 Am M\"uhlenberg 1, 14476 Golm, Germany\\

\end{center}
\vspace{20pt}
\begin{center}
Abstract
\end{center}
\noindent
Starting from 2D Euclidean quantum gravity, we show 
that one recovers 2D Lorentzian quantum gravity by 
removing all baby universes. Using a peeling procedure to
decompose the discrete, triangulated geometries
along a one-dimensional path, we explicitly associate with
each Euclidean space-time a (generalized) Lorentzian space-time.
This motivates a map between the parameter spaces of the two
theories, under which their propagators get identified. In two dimensions,
Lorentzian quantum gravity can therefore be viewed as a ``renormalized'' 
version of Euclidean quantum gravity.

\vfill{\noindent PACS codes: 02.10.Eb, 04.20Gz, 04.60.Nc, 05.20.y\\
Keywords: 2D gravity, random triangulations, Lorentzian triangulations,
transfer matrix formalism, random walk, branched polymers}
\newpage

\pagestyle{plain}
\setcounter{page}{1}

\section{Introduction}
The role of baby universes in quantum gravity has been discussed 
extensively and many conjectures have been made concerning their 
role in providing effective coupling constants for field theories 
as well as for gravity itself~\cite{babyuniverses}. 
Due to the lack of a theory of (four-dimensional) quantum gravity,
the ``integration over baby universes'' has never been performed 
in any explicit manner.
However, there is a chance that this idea can be realized
in two dimensions, where quantum gravity is a renormalizable quantum 
field theory.
In this article we will show that starting from 2D Euclidean 
quantum gravity, there is indeed a sense in which one can integrate 
out all baby universes.
The resulting ``renormalized'' quantum theory coincides with 
the theory of so-called Lorentzian 2D quantum gravity.

The Lorentzian quantum gravity model was introduced in order 
to have a path-integral formulation where the Lorentzian character
of the metric is built in at a fundamental level. 
It is also closer to canonical quantization methods, where 
from the outset one works with globally hyperbolic space-times. 
The two-dimensional model can be solved explicitly  
and is different from Euclidean quantum gravity, by which 
we will mean the theory defined by a path integral over
Euclidean geometries.
In the continuum limit, a typical geometry 
coming from the path integral of Lorentzian quantum gravity is 
two-dimensional~\cite{Ambjorn:1998xu,Ambjorn:1998fd}. 
This is in contrast with the situation 
in Euclidean quantum gravity, where a typical geometry 
has an anomalous fractal Hausdorff dimension of 
four~\cite{AW}.

We will be using a representation of Euclidean quantum gravity 
in terms of so-called dynamical triangulations. 
This method provides a reparameterization-invariant 
regularization of the field theory. The two-dimensional geometries 
will be represented by equilateral triangulations 
(with edges of cut-off length $a$). 
The path integral is performed by summing over all 
triangulations of a given topology. The continuum limit is obtained 
as the lattice spacing $a \to 0$. 
We will then give an explicit prescription for removing baby universes
from a given Euclidean triangulation.
One may view this procedure 
as analogous to the removal of tadpole graphs in a quantum field 
theory. The latter can usually be viewed as a redefinition 
of the coupling constants and a shift of the fields in the theory.
In our case the effect is more drastic.
We end up with a new, physically inequivalent theory.

\section{The two-loop correlator and the peeling procedure}

\subsection{The Euclidean case} 

We will investigate the relation between two-dimensional
Euclidean and Lorentzian 
quantum gravity by studying the distance-dependent two-loop
correlator of the two models. 
The distance between two links $l$ and $l'$ in a given
triangulation $T$ is defined as
the length of the shortest path connecting $l$ and $l'$, 
taken along the links of the dual lattice.
Furthermore, the distance between a link $l$
and a loop $\cal L$ is defined as the minimum distance of the link
$l$ to a link belonging to ${\cal L}$. With these definitions a loop
${\cal L}_2$ is said to have distance $t$ to a loop ${\cal L}_1$ if all
links in ${\cal L}_2$ have distance $t$ to the loop ${\cal L}_1$.

The distance-dependent two-loop correlation function $G(l_1,l_2,t)$
is defined as
\beq
G(l_1,l_2,t)=\sum_{T\in {\cal T} (l_1,l_2,t)}g^{N(T)},
\label{def1}
\eeq
where the sum goes over all triangulations of cylindrical topology with an
exit loop ${\cal L}_2$ of length $l_2$ and an entrance loop
${\cal L}_1$ of length $l_1$ containing a marked link, the distance of 
${\cal L}_2$ from ${\cal L}_1$ being equal to $t$. The
quantity $N(T)$ is the number of triangles in the triangulation $T$
and $g$ is related to the cosmological constant $\mu$ by $g=e^{-\mu}$.

Euclidean and Lorentzian quantum gravity differ by the type
of triangulations allowed in~\rf{def1}. In the Lorentzian
case one only sums over triangulations which for any
$t'\leq t$ have only one (connected) loop at distance
$t'$ from the entrance loop. In the usual language of 2D quantum
gravity this means that baby-universes are forbidden. Furthermore, in
the Lorentzian case a loop cannot shrink to length zero.
Using the distinguished variable $t$ as a time variable, we can 
divide the links into space-like and (future-oriented) time-like,
and equip every Lorentzian
triangulation with a well-defined causal structure~\cite{Ambjorn:1998xu}. 
By contrast, in the Euclidean
case there can be more than one loop at a given distance $t$ from
the entrance loop and the above construction does not work. 
We note, however, that in~\cite{BJ} a suggestion
of introducing a notion of space-like and time-like edges for Euclidean
triangulations was put forward. This lead 
to a model equivalent to 2D Euclidean quantum
gravity coupled to Ising spins.

It is well-known that a combinatorial identity can be derived for the
Euclidean version of~\rf{def1} 
by a so-called peeling procedure~\cite{Watabiki:1995ym}. 
We will show that also in the Lorentzian case a peeling procedure 
can be defined, which in addition enables us to set up a direct
correspondence between the discrete Euclidean and Lorentzian histories.

We start by reviewing briefly the peeling formalism in the Euclidean
case. In order to specify a peeling, one needs a set of moves that
systematically remove building blocks at a marked link on the 
boundary of a triangulation and a rule for choosing a new marked link
on the boundary after each move. 
We will work in the class of so-called unrestricted Euclidean
triangulations which apart from triangles also contain double
links as building blocks. (Unrestricted triangulations constitute the class
of triangulations appearing in the usual matrix model formulation of 2D
gravity.) This generalization does not affect any of the universal
properties of the model.
Consider now an unrestricted triangulation contributing to the sum
in~\rf{def1}. By definition it has an entrance loop with one marked link, 
which belongs either to a triangle or a
double link. If it belongs to a triangle we remove the triangle
from the triangulation. We distinguish three cases, according to
whether the triangle has one, two or three links in common with the
boundary. 
If it has more than one boundary link, additional double links
are created during the move, 
so that the length of the original boundary loop
is always increased by one (see Fig.1, where also the rule for link
marking is specified). 
\begin{figure}[htb]
\centerline{\epsfxsize=6.truecm\epsfbox{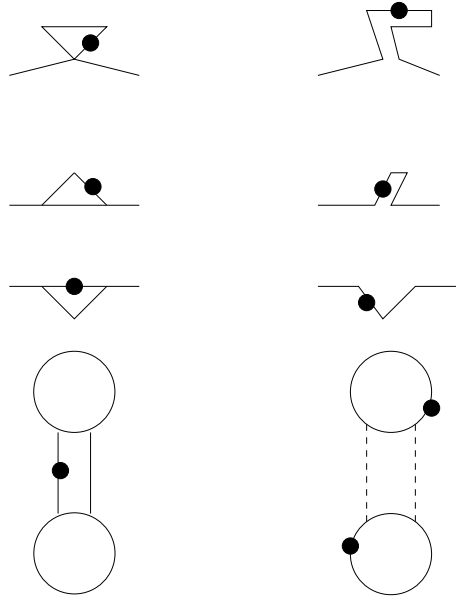}}
\caption{The minimal decomposition. In the left column the
boundary before the decomposition and in the right column the
boundary after the decomposition.}
\end{figure}
If the marked link belongs to a double link we simply remove the
double link. This will in general lead to a
splitting of the original loop into two. 
On each of the resulting loops we place a mark as shown in Fig.~1. We
can choose either of the two loops as our new entrance loop,
associating the other one with a baby universe. This gives rise to the
factor of two in eqn.~\rf{diffl} below. (One or both loops may of course be
trivial, i.e. consist of only ``a marked point''.)
By this procedure we associate with each Euclidean triangulation
a unique sequence of peeling moves.
If the entrance loop has length $l_1$, then on average $l_1$
peeling steps will move the boundary one time step ahead.
Thus one identifies the deformation $\delta G(l_1,l_2,t)$ associated with
a single peeling step with $\displaystyle{\frac{1}{l_1} 
\frac{\partial}{\partial t} G(l_1,l_2,t)}$. 
This deformation fulfils
\beq
\frac{\partial}{\partial t}G(l_1,l_2,t)=-l_1G(l_1,l_2,t)+
gl_1 G(l_1+1,l_2,t)+
2\sum_{l=0}^{l_1-2} l_1 W(l)G(l_1-l-2,l_2,t),
\label{diffl}
\eeq
with the initial condition
\beq
G(l_1,l_2,t=0)=\delta_{l_1,l_2},
\eeq
where $W(l)$ is the disk amplitude defined by
\beq
W(l)=\sum_{T\in {\cal T}(l)} g^{N(T)},
\label{Wl}
\eeq
and the sum is over all (non-restricted) triangulations with one connected
boundary component of length $l$. Let us introduce the generating functionals
\bea
G(z,y,t)&=&\sum_{l_1,l_2=0}^{\infty} \frac{1}{z^{l_1+1}}\frac{1}{y^{l_2+1}}\,
G(l_1,l_2,t),
\label{Gz} \\
W(z)&=&\sum_{l=0}^{\infty}\frac{1}{z^{l+1}}\, W(l).
\label{W0z}
\eea 
Here $z$ and $y$ can be understood as boundary cosmological constants. Then
the differential equation~\rf{diffl} reads
\beq
\frac{\partial}{\partial t} G(z,y,t)=\frac{\partial}{\partial z}
\left(h(z)G(z,y,t)\right), \label{diffx}
\eeq
with initial condition
\beq
\hspace{0.7cm}G(z,y,t=0)=\frac{1}{zy}\frac{1}{zy-1},
\eeq
where
\beq
h(z)=z-gz^2-2 W(z).
\label{h}
\eeq
It is well known that 
in analogy with the procedure described above,
it is possible to derive a combinatorial identity for 
$W(z)$ by removing a
triangle or a double link from a triangulation contributing to the sum
in~\rf{Wl}, see for instance~\cite{Ambjorn:1997di}. 
The combinatorial identity reads
\beq
(z-gz^2)W(z)-1+g(w_1(g)+z)=\left(W(z)\right)^2,
\eeq  
where
\beq
w_1(g)=\sum_{T\in {\cal T}(1)} g^{N(T)}.
\eeq
Here the terms proportional to $g$ emerge when a triangle is removed whereas
the term $(W(z))^2$ emerges when a double link is removed and the surface
splits into two.
We note that if we set $g=0$ we get
\beq
W(z)=\frac{1}{2}\left(z-\sqrt{z^2-4}\right),
\eeq
which is the well-known generating function for rooted branched polymers. These
appear exactly because of our inclusion of double links.
For $g\neq 0$, $W(z)$ is the solution of a quadratic equation and the unknown
constant $w_1(g)$ is determined by noticing that the analyticity structure of
$W(z)$ should not change discontinuously at $g=0$. The solution for $W(z)$
reads
\beq
W(z)=\frac{1}{2}\left(z-gz^2+g\left(z-c(g)\right)\sqrt{(z-c_+(g))(z-c_-(g))}
\right),
\eeq
where $c_-(g)\leq c_+(g)\leq c(g)$ and where all three quantities can be
expressed explicitly in terms of $g$. At a certain value $g_c$ of $g$ the
radius of convergence of the series~\rf{Wl} is reached and a continuum limit
can be defined. Setting 
\beq
g_c-g=a^2\Lambda,
\label{euscal1} 
\eeq
where $a^2$ is a scaling parameter
with the dimension of area and $\Lambda$ is the continuum cosmological
constant we have for $g$ close to $g_c$
\bea
c(g)&=&z_c\left(1+\frac{1}{2}\,a\sqrt{\Lambda}\right)
+{\cal O}(a^2),\\
c_+(g)&=&z_c\left(1-a\sqrt{\Lambda}\right)+{\cal O}(a^2),\\
c_-(g)&=&c_-+{\cal O}(a^2). 
\eea
It is now natural to introduce a continuum boundary cosmological constant $Z$
by
\beq
z=z_c\left(1+aZ\right).
\label{euscal2}
\eeq
We thus have that $h(z)$ exactly consists of the scaling part of $-2W(z)$.
Introducing a continuum time variable $T$ by
\beq
T=z_c^{1/2}(z_c-c_-)^{1/2}\, g_c\, a^{1/2} t,
\eeq
one obtains a differential equation for the continuum version
$G(Z,Y,T)$ of the two-loop correlator, namely,
\beq
\frac{\partial}{\partial T} G(Z,Y,T) =
-\frac{\partial}{\partial Z}
\left(\left((Z-\frac{1}{2}\sqrt{\Lambda}) \sqrt{Z+\sqrt{\Lambda}}\right)\,
G(Z,Y,T)\right),
\label{Euclid}
\eeq
with the initial condition
\beq
G(Z,Y,T)=\frac{1}{Z+Y}.
\label{Eubound}
\eeq
Note that the continuum cosmological constants $\Lambda$ and $Z$ and
the time $T$ are only defined up to positive multiplicative constants.
We have used this freedom to bring the propagator equation into the
simple form (\ref{Euclid}).

\subsection{The Lorentzian case} 

One may invert the peeling procedure introduced in the previous
section to obtain a Euclidean triangulation
from a sequence of peeling moves. Its geometry is given in the form of a bulk
cylindrical geometry connecting the entrance and exit loops,
``decorated'' with baby universe out-growths.
This is the case because some of the peeling steps remove just a single
triangle (which we now associate with the bulk cylinder), while the
removal of a double link in general amounts to the deletion
of an entire baby universe. 

As mentioned earlier, the difference between Euclidean and
Lorentzian gravity can be traced to the presence or absence of
baby universes. There is already an explicit construction of Euclidean
from Lorentzian histories through the addition of baby universes 
\cite{Ambjorn:1998xu}. Our current aim is to go the other way. 
The peeling procedure suggests an obvious map from Euclidean
to Lorentzian discrete geometries. 
Given a representation of
a Euclidean geometry as a sequence of peeling moves, consider a step
where a double link-cum-baby universe is removed, see Fig.~2.
\begin{figure}[htb]
\centerline{\epsfxsize=6.truecm\epsfbox{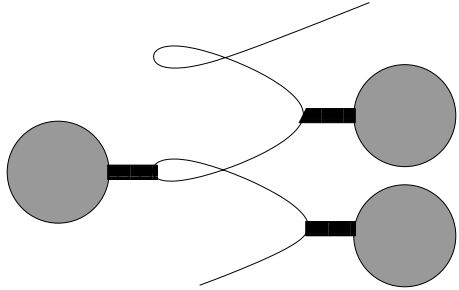}}
\caption{A representation of a Euclidean triangulation as a sequence
of peeling moves along a ``spiral-like'' one-dimensional curve.
Black rectangles are double links and grey circles baby universes.}
\end{figure}
Decompose the baby universe into a branched polymer attached to the 
marked link and a smaller baby universe attached to the branched
polymer in such a way that the branched polymer consists of the
largest possible number of links. Next, replace the step where the
double link-cum-baby universe is removed by a step where only the double
link and the branched polymer is taken way. Do the same for all
similar steps in the sequence of peeling moves. The resulting sequence
of peeling moves obviously represents a geometry without baby
universes. It is clear that the class of geometries one generates this
way is larger than the class of Lorentzian triangulations introduced
in~{\cite{Ambjorn:1998xu}. Firstly, it
contains all of the regular Lorentzian triangulations of
\cite{Ambjorn:1998xu}, as can be checked explicitly. Secondly,
the boundaries may now contain double links. 
We will refer to this generalized class of geometries as
``Lorentzian geometries''. This will be justified below by
showing that they lead to a theory in the same universality class
as the original Lorentzian model and the extensions discussed
in \cite{DiFrancesco:1999em}. The differential equation for the
two-loop correlator of Lorentzian geometries is obtained by replacing 
$2W(z)$  in~\rf{h} by $W_{BP}(z)$,  and thus reads
\beq
\frac{\partial}{\partial \tilde{t}}
\tilde{G}(\tilde{z},\tilde{y},\tilde{t})
=\frac{\partial}{\partial \tilde{z}}\left(\tilde{h}(\tilde{z})
 \tilde{G}(\tilde{z},\tilde{y},\tilde{t})\right),
\eeq
with initial condition
\beq
\tilde{G}(\tilde{z},\tilde{y},\tilde{t}=0)=\frac{1}{\tilde{z}\tilde{y}}
\frac{1}{1-\tilde{z}\tilde{y}},
\eeq
where
\beq
\tilde{h}(\tilde{z})=\tilde{z}-\tilde{g}\tilde{z}^2-W_{BP}(\tilde{z}).
\label{htilde}
\eeq
We have introduced tildes on all the Lorentzian variables in
order to distinguish them from the Euclidean ones Let us note here that it
is easy to generalize this model of Lorentzian quantum gravity to models 
with arbitrary polygons as building blocks. Using the newly developed
technology of~\cite{HW} it may be possible to solve all of these models
exactly. Here we only consider the simplest model.  
We wish to
locate a critical point $(\tilde{g}_c,\tilde{z}_c)$ at which we can
define a non-trivial continuum limit by the scaling
relations
\beq
\tilde{g}=\tilde{g}_c(1-c_1  a^2\tL),\hspace{0.7cm} 
\tilde{z}=\tilde{z}_c\left(1+ c_2 a\tilde{Z}\right),
\label{loscal}
\eeq
where $c_1$ and $c_2$ are numerical constants.
The only way to obtain a non-trivial differential
equation in the continuum limit is by demanding that
$\tilde{h}(\tilde{z})\sim {\cal O}(a^2)$. This determines the critical
values of $\tilde{z}$ and $\tilde{g}$ to be
\beq
\tilde{z}_c=\frac{4}{\sqrt{3}},\hspace{0.7cm}
\tilde{g}_c=\frac{3\sqrt{3}}{16}.
\label{crit}
\eeq
For later convenience we set $c_2=\frac{1}{3^{3/4}}$ and
$c_1=\frac{1}{{\sqrt{3}}}$. We also introduce a continuum version
$\tilde{T}$ of $\tilde t$ by
\beq
\tilde{t}=\frac{1}{a}\,\tilde{T}.
\eeq
Note that the relative dimensions of the time variable and 
the area are different for Euclidean and Lorentzian
quantum gravity. Then we derive 
the following evolution equation 
for the continuum two-loop correlator
$\tilde{G}(\tilde{Z},\tilde{Y},\tilde{T})$
\beq
\frac{\partial}{\partial \tilde{T}}\tilde{G}(\tilde{Z},\tilde{Y},\tilde{T})
=-\frac{\partial}{\partial \tilde{Z}}\left(\left(\tilde{Z}^2-\tL\right)
\tilde{G}(\tilde{Z},\tilde{Y},\tilde{T})\right),
\label{Lorentz}
\eeq
with initial condition
\beq
\tilde{G}(\tilde{Z},\tilde{Y},\tilde{T})=\frac{1}{\tilde{Z}+\tilde{Y}}.
\label{Lobound}
\eeq
This differential equation coincides with the one derived
in~\cite{Ambjorn:1998xu}, using a completely different strategy.

Let us now turn to a discussion of the disk amplitude $\tilde{W}(l)$
of Lorentzian gravity.
At the discrete level it may be defined as the sum
$\sum_{\tilde{t}=1}^{\infty}\tilde{G}(l,l_2=0,\tilde{t})$.
This reflects the fact
that a Lorentzian geometry with a regular initial spatial geometry 
at $\tilde{t}=0$ can only terminate at a finite $\tilde{t}$ if we allow for a 
singularity where space contracts to a point. 

We can derive a combinatorial identity for the generating functional 
$\tilde{W}(\tilde{z})$ in the same
way as for the Euclidean disk amplitude $W(z)$. The combinatorial
identity reads  
\beq
(\tilde{z}-\tilde{g}\tilde{z}^2)\tilde{W}(\tilde{z})-1+
\tilde{g}(\tilde{w}_1(\tilde{g})+\tilde{z})=
W_{BP}(\tilde{z})\tilde{W}(\tilde{z}).
\eeq
As in the case of the two-loop correlator, 
the only modification of the equation
concerns the term associated with the removal of a double link. If, in
the Euclidean case, the marked link is part of a double link, we remove
the double link and are left with two disjoint disks. If, in the
Lorentzian case, the marked link is part of a double link, we remove 
the double link and are left with one disk and one branched polymer.
In this case, 
by setting $\tilde{g}=0$, we get
\beq
\tilde{W}(\tilde{z})=W_{BP}(\tilde{z}).
\label{wtilde}
\eeq 
This
underlines the consistency of our construction. For
$\tilde{g}\neq 0$ on the other hand we
obtain 
\beq
\tilde{W}(\tilde{z})=\frac{2(1-\tilde{g}
(\tilde{w}_1(\tilde{g})+\tilde{z}))}
{\tilde{z}-2\tilde{g}\tilde{z}^2+\sqrt{\tilde{z}^2-4}}.
\label{wtildeg}
\eeq
Note that the denominator of~\rf{wtildeg} is equal to
$2\tilde{h}(\tilde{z})$. For $\tg<\frac{1}{4}$ this function has one zero
$\tilde{z}_1(\tilde{g})$, which behaves as
$\frac{1}{\tilde{g}}+{\cal O}(\tilde{g})$ when $\tilde{g}$ is close to zero.
Hence a pole appears 
in the transition from~\rf{wtilde} to~\rf{wtildeg} and
we must choose the unknown constant $\tilde{w}_1(\tilde{g})$ 
such that this pole is cancelled. 
At $\tg=\frac{1}{4}$ the function $\tilde{h}(\tz)$ acquires an additional 
zero, $\tilde{z}_2(\tilde{g})$,
and the critical point
$\tilde{g}_c=\frac{3\sqrt{3}}{16}$ corresponds to the situation where the
two zeros $\tilde{z}_1(\tg)$ and $\tz_2(\tg)$ coincide. 
In the vicinity of the critical point 
we have the expansions
\bea
\tilde{z}_1(\tg)&=&\frac{4}{\sqrt{3}}+\frac{16}{3^{7/4}}
(\tilde{g}_c-\tilde{g})^{1/2},\label{scalz1} \\
\tilde{z}_2(\tg)&=&\frac{4}{\sqrt{3}}-\frac{16}{3^{7/4}}
(\tilde{g}_c-\tilde{g})^{1/2}.\label{scalz2}
\eea
Using~\rf{scalz1},~\rf{scalz2} and~\rf{loscal}, we
derive the continuum disk amplitude
\beq
\tilde{W}(\tilde{Z})=\lim_{a\rightarrow 0}\,\, a \,\tilde{W}(\tz)=
\frac{4}{3^{5/4}}\frac{1}{\tilde{Z}+\sqrt{\tL}}.
\label{Ldisk}
\eeq
This expression coincides with the one found in~\cite{Ambjorn:1998xu} 
by completely different means.

\section{Renormalizing Euclidean gravity to
Lorentzian gravity}\label{map}

We will now show that it is possible to ``renormalize'' the 
Euclidean coupling constants and the Euclidean time such that 
the equation governing the time evolution of the Euclidean two-loop 
function is mapped to the corresponding equation for the Lorentzian 
two-loop function.

We start from
the discrete version of the differential equation for the
time development of the Euclidean universes,~\rf{diffx}, and 
make an ansatz
\beq\label{junk1}
\tilde{z}=\tilde{z}(z,g),\hspace{0.7cm} \tilde{t}=\frac{1}{b} t, 
\hspace{0.7cm} \tilde{g}=\tilde{g}(g),
\eeq
where $b$ may depend on $g$ and $\tilde{g}$, but not on $z$ or $\tilde{z}$.
In principle it should be possible to determine the required map
between the coupling constants of the two theories 
directly from the peeling procedure which maps Euclidean to
Lorentzian histories. However, it turns out to be simpler to use an
indirect argument. 
{\it If} there exists a transformation of the form \rf{junk1} which 
maps the Euclidean differential equation into the Lorentzian differential 
equation, then clearly it must  include a wave-function renormalization
of the two-loop correlator
\beq
\tilde{G}(\tz,\ty,\tilde{t})\propto 
\left(\frac{\partial\tilde{z}}{\partial z}\right)^{-1}
G(z(\tz ),y(\ty ),b\tilde{t}),
\label{waveren}
\eeq
and furthermore
\beq\label{junk2}
b\, h(z)\frac{\partial \tilde{z}}{\partial z} =\tilde{h}(\tilde{z}),
\eeq
or, equivalently,
\beq\label{integral}
\int \frac{dz}{h(z)}=b \int \frac{d\tilde{z}}{\tilde{h}(\tilde{z})}+k,
\eeq
where $k$ is an integration constant.

Using the explicit expressions~\rf{h} and \rf{htilde} for $h(z)$ and
$\tilde{h}(\tilde{z})$, we can carry out the integrations in~\rf{integral}
and find
\begin{eqnarray}
\lefteqn{\hspace{-0.1cm}
-\frac{1}{\sqrt{\delta(g)}}\log\left\{
\frac{2\delta(g)+\epsilon(g)(z-c(g))
+2\sqrt{\delta(g)}\sqrt{(z-c_+(g))(z-c_-(g))}}
{z-c(g)}\right\}} \label{discretemap} \\
&&\hspace{-0.5cm}=b\left\{\frac{1}{2\tg^2}\left[
\frac{1-2\tg \tilde{z}_1(\tilde{g})}
{2\tilde{z}_1(\tilde{g})\left(\tz_1(\tg)-\frac{3}{4\tg}\right)} 
\log(\tilde{z}-\tilde{z}_1(\tilde{g}))+
\frac{1-2\tg \tilde{z}_2(\tilde{g})}
{2\tilde{z}_2(\tilde{g})\left(\tz_2(\tg)-\frac{3}{4\tg}\right)} 
\log(\tilde{z}-\tilde{z}_2(\tilde{g}))
\right.\right.
\nonumber\\
&&\left.\left.
+\frac{\sqrt{\tz_1(\tg)^2-4}}{4\tz_1(\tg)^2
\left(\tz_1(\tg)-\frac{3}{4\tg}\right)}
\log\left[\sqrt{\tz_1(\tg)^2-4}\sqrt{\tz^2-4}+\tz\tz_1(\tg)-4\right]+\ldots
\right]\right\}+k,
\nonumber
\end{eqnarray}
Here 
$\ldots$ means terms similar to the previous one
with $\tz_1(\tg)$ replaced by $\tz_2(\tg)$, $\tz_3(\tg)$ and $\tz_4(\tg)$
where $\tz_1(\tg)$, $\tz_2(\tg)$, $\tz_3(\tg)$ and $\tz_4(\tg)$ are the
four roots of the polynomial $z^4-\frac{1}{\tg}\,z^3+\frac{1}{\tg^2}$ 
with $\tz_1(\tg)$ and $\tz_2(\tg)$ still playing the roles 
described on the previous page.
Furthermore $\delta$ and $\epsilon$ are defined by  
\bea
\delta(g)&=&(c(g)-c_+(g))(c(g)-c_-(g)), \nonumber \\
\epsilon(g)&=& 2c(g)-c_+(g)-c_-(g).
\nonumber
\eea
For any value of $k$, $g$, $\tilde{g}$ 
and $b$, the relation \rf{discretemap} defines a mapping
$\tilde{z}=\tilde{z}(z)$. However, not all values of $b$ and $k$ correspond  
to situations we can encounter in the peeling procedure. Firstly, 
the two theories have to approach criticality simultaneously, since 
macroscopic surfaces and boundaries 
are created at the critical points. Secondly, we insist on a canonical 
scaling dimension of the cosmological and boundary cosmological constants.
This amounts to saying that the scaling of the continuum cosmological 
constants is governed by the same power of the lattice cut-off $a$.
Since we have assumed that $\tilde{g}$ is only a function of $g$, 
and not of $z$ and $\tilde{z}$, this implies that $\tL \propto \L$, with a
proportionality coefficient independent of the cut-off.
In the scaling limit all dependence on $\tz_3(\tg)$ and $\tz_4(\tg)$
disappears and using the scaling relations~\rf{euscal1}--\rf{euscal2},
\rf{loscal}, \rf{scalz1} and \rf{scalz2}, we see that in
order for the mapping to persist in the continuum limit, 
$b$ and $k$ must scale according to
\beq\label{singular}
b=a^{1/2}\Lambda^{1/4} \beta, \hspace{0.7cm} 
k=a^{-1/2}\Lambda^{-1/4}
\kappa,
\eeq
for two dimensionless constants $\beta$ and $\kappa$.
To fix the relevant power of $\Lambda$ in \rf{singular},
we have used the fact that $b$ and $k$ cannot 
depend on $z$ or $\tilde{z}$.
{\it Note that $b$ and $k$ behave 
in a singular way when we approach the continuum limit}.
This is an unavoidable consequence of interpolating
between two universality classes of 2D geometries with
different scaling behaviour (recall that  
in the previous section we showed that $t$ scales as $t\sim a^{-1/2}$, 
whereas $\tilde{t}\sim a^{-1}$).

It is now straightforward to compute the continuum version of 
\rf{discretemap},
\beq
\frac{1}{\SL}\log\left\{
\frac{\sqrt{\frac{2}{3}}\Lambda^{1/4}\sqrt{Z+\sqrt{\Lambda}}+\sqrt{\Lambda}}
{\sqrt{\frac{2}{3}}\Lambda^{1/4}\sqrt{Z+\sqrt{\Lambda}}-\sqrt{\Lambda}}\right\}
=\frac{\beta}{\tSL}  \log\left\{
\frac{\tilde{Z}+\sqrt{\tL}}{\tilde{Z}-\sqrt{\tL}}\right\}
+\frac{\kappa}{\sqrt{\Lambda}}, 
\label{ztildezcont}
\eeq
where we have absorbed some additional numerical constants into
$\beta$ and $\kappa$. 
We will argue that in order for this map to be associated 
with the cutting procedure we must choose $\k \equ 0$.  
Taking the limit $Z \to \infty$ implies that the boundary length 
$L \to 0$. Now, a Lorentzian boundary is shorter than a
Euclidean one, since it was obtained from the latter by cutting away
baby universes which at a given time contributed to the
boundary length. 
This is the physical motivation for requiring that $Z\to \infty$ should 
imply $\tZ \to \infty$, i.e.\ $\k \equ 0$. Moreover, a rescaling of 
both $Z$ and $\SL$ with a common factor allows us to choose 
\beq\label{3.14a}
\SL =  \frac{1}{\b} {\tSL},
\eeq
in such a way that the map \rf{ztildezcont} simplifies to 
\beq\label{3.14}
\frac{\tZ}{\tSL} = \sqrt{\frac{2}{3}}\; \frac{\sqrt{Z +\SL}}{\FL},
\eeq
and the relation between the continuum times $T$ and $\tT$ becomes
\beq\label{3.12}
\tT= \frac{T}{\b \FL}.
\eeq
One can of course verify directly that the (non-analytic) change 
of variables \rf{3.14} and \rf{3.12} supplemented by the wave function
renormalization~\rf{waveren}
transforms eq.\ \rf{Euclid} 
for Euclidean quantum gravity into eq.\ \rf{Lorentz} for Lorentzian 
gravity.\footnote{Note that some additional constants have been absorbed
into $\b$ in going from~\rf{discretemap} to~\rf{ztildezcont}}. 
To map the actual two-loop correlator of Euclidean quantum gravity
$G_0(Z,Y,T)$ (i.e.\ the solution to~\rf{Euclid} with initial 
condition~\rf{Eubound}) onto the actual two-loop correlator of
Lorentzian quantum gravity $\tilde{G}_0(\tilde{X},\tilde{Y},\tilde{T})$
(i.e.\ the solution to~\rf{Lorentz} with initial condition~\rf{Lobound}) 
an additional dressing factor is needed
More precisely one has
\beq\label{3.15}
\tG_0(\tZ,\tY,\tT) = K(\bZ(T,Z),Y)\; G_0(Z,Y,T),
\eeq
where
\beq\label{3.16}
K(\bZ(T,Z),Y) = \frac{\bZ(T,Z)+Y}{\tZ(\bZ(T,Z))+\tY(Y)}
\, \frac{\sqrt{Z+\SL}}{\sqrt{\bZ(T,Z)+\SL}},
\eeq
and where $\bZ(T,Z)$ is the solution to characteristic equation
for \rf{Euclid}:
\beq\label{3.2}
\frac{\pa \bZ(T,Z)}{\pa T} = -(Z-\oh \SL)\sqrt{Z+\SL},~~~~\bZ(T\equ 0,Z)= Z.
\eeq
Summing up, we have shown that ``integrating over baby'' universes can
be described as a renormalization of the cosmological constants and
the time variable combined with a dressing of the two-loop correlator.

\section{Discussion}

We have shown how one can analyze the geometric structure of Lorentzian
and Euclidean triangulations by representing them as a sequence of
peeling moves defining a one-dimensional `spiral-like' path. 
Along this 
path, one meets in both cases either triangles or double links. 
The
difference between the Euclidean and Lorentzian cases is that in the
former a double link may have a branched polymer plus
an entire baby universe attached
to it whereas in the latter a double link can have at most a branched
polymer attached to it.
At the level of sequences of peeling moves, this allows us to define a  
many-to-one map from Euclidean to Lorentzian geometries.
This is reminiscent of the situation one encounters in
analyzing the fractal structure of random walks versus branched
polymers. There, moving along a path connecting two vertices, in the
random walk case one may have only a single link emerging at
each vertex. For the branched polymer case, there can be an entire 
polymer attached at each vertex, see e.g.~\cite{Ambjorn:1997di}. 
In this context it may be of interest to point out that
pure Lorentzian quantum gravity has been proven to be equivalent to
a model of random walks~\cite{DiFrancesco:1999em}.

This analogy is further strengthened by the nature of the continuum 
renormalization of the coupling constants
presented in section~\ref{map}. Apart from trivial 
rescalings, only the {\it boundary}
cosmological constant is subject to a renormalization. This reflects the
fact that the peeling decomposition proceeds along a one-dimensional
path. Moreover, the functional form of the relation between the
Lorentzian and Euclidean boundary cosmological constants is very
similar to that between the cosmological constants of random walks 
and branched polymers, since in both cases the cosmological constant of 
the simpler 
model has a square-root dependence on the cosmological 
constant of the more extended
model. Furthermore, in both cases the fractal dimension of the
decorated system is twice that of the simpler system.


\section*{Acknowledgements}
J.A. acknowledges the support of MaPhySto, Centre for Mathematical Physics,
funded by the Danish National Research Foundation.
J.C. acknowledges the support of the EU via grant ERB
4001GT973188. C.K.\ acknowledges the support of the Carlsberg Foundation.


\end{document}